\documentclass[aps,prl,preprint,nopacs,superscriptaddress]{revtex4}
\usepackage{graphicx}
\usepackage{verbatim}
\usepackage{mathrsfs}
\usepackage{amsmath,amsfonts,amssymb}
\usepackage{graphicx}

\def\3{2.8in}    
\def\2{2.5in}
\def\4{3.0in}

\def \beq {\begin{equation}}
\def \eeq {\end{equation}}
\begin{document}

\title{Momentum space imaging of Cooper pairing in a half-Dirac-gas topological superconductor (a helical 2D topological superconductor)}
\author{Su-Yang Xu}\affiliation {Joseph Henry Laboratory, Department of Physics, Princeton University, Princeton, New Jersey 08544, USA}
\author{Nasser Alidoust}\affiliation {Joseph Henry Laboratory, Department of Physics, Princeton University, Princeton, New Jersey 08544, USA}
\author{Ilya Belopolski}\affiliation {Joseph Henry Laboratory, Department of Physics, Princeton University, Princeton, New Jersey 08544, USA}
\author{Anthony Richardella}\affiliation {Department of Physics, The Pennsylvania State University, University Park, Pennsylvania 16802-6300, USA}
\author{Chang Liu}\affiliation {Joseph Henry Laboratory, Department of Physics, Princeton University, Princeton, New Jersey 08544, USA}
\author{Madhab Neupane}\affiliation {Joseph Henry Laboratory, Department of Physics, Princeton University, Princeton, New Jersey 08544, USA}
\author{Guang Bian}\affiliation {Joseph Henry Laboratory, Department of Physics, Princeton University, Princeton, New Jersey 08544, USA}

\author{Song-Hsun Huang} \affiliation{Center for Condensed Matter Sciences, National Taiwan University, Taipei 10617, Taiwan}
\author{Raman Sankar} \affiliation{Center for Condensed Matter Sciences, National Taiwan University, Taipei 10617, Taiwan}

\author{Chen Fang}\affiliation {Joseph Henry Laboratory, Department of Physics, Princeton University, Princeton, New Jersey 08544, USA}\affiliation {Department of Electrical and Computer Engineering, University of Illinois, Urbana, Illinois 61801, USA}
\author{Brian Dellabetta}\affiliation {Department of Electrical and Computer Engineering, University of Illinois, Urbana, Illinois 61801, USA}

\author{Wenqing Dai}\affiliation {Department of Physics, The Pennsylvania State University, University Park, Pennsylvania 16802-6300, USA}
\author{Qi Li}\affiliation {Department of Physics, The Pennsylvania State University, University Park, Pennsylvania 16802-6300, USA}
\author{Matthew J. Gilbert}\affiliation {Department of Electrical and Computer Engineering, University of Illinois, Urbana, Illinois 61801, USA}
\author{Fangcheng Chou} \affiliation{Center for Condensed Matter Sciences, National Taiwan University, Taipei 10617, Taiwan}
\author{Nitin Samarth}\affiliation {Department of Physics, The Pennsylvania State University, University Park, Pennsylvania 16802-6300, USA}
\author{M. Zahid Hasan}\affiliation {Joseph Henry Laboratory, Department of Physics, Princeton University, Princeton, New Jersey 08544, USA}


\begin{abstract}

Superconductivity in Dirac electrons has recently been proposed as a new platform between novel concepts in high-energy and condensed matter physics. It has been proposed that supersymmetry and Majorana fermions, both of which remain elusive in particle physics, may be realized as emergent particles in superconducting Dirac electron systems. Using artificially fabricated topological insulator-superconductor heterostructures, we present direct spectroscopic evidence for the existence of Cooper pairing in a half Dirac gas 2D topological superconductor. Our studies reveal that superconductivity in a helical Dirac gas is distinctly different from that of in an ordinary two-dimensional superconductor while considering the spin degrees of freedom of electrons. We further show that the pairing of Dirac electrons can be suppressed by time-reversal symmetry breaking impurities removing the distinction. Our demonstration and momentum-space imaging of Cooper pairing in a half Dirac gas and its magnetic behavior taken together serve as a critically important 2D topological superconductor platform for future testing of novel fundamental physics predictions such as emergent supersymmetry and quantum criticality in topological systems.

\end{abstract}
\date{\today}
\maketitle

Realization of novel superconductivity is one of the central themes in condensed matter physics in general \cite{TI_book_2014, SUSY, Chubukov, Kane_Proximity, Zhang_TSC, Patrick Lee, Sarma, Hosur, RMP, Zhang_RMP, Hor, Mason, Morpurgo, Nitin, Kanigel, Dong_PRL, LuLi1, Gordon, Leo, Ando, Patrick Lee2, TeWari, Marcus, Franceschi}. Superconductivity is a collective phenomenon, where electrons moving to the opposite directions (${\pm}k$) form dynamically bound pairs, resulting in a Cooper pair gas. In an ordinary superconductor, the conduction electrons that move along a certain direction have both spin-up and spin down electrons available for the Cooper pairing. The superconductivity observed so far, including in the conventional $s$-wave BCS superconductors as well as the cuprate or heavy fermion $d$-wave superconductors, all share this property. Recently, the discovery of 3D topological insulator in bismuth based semiconducting compounds have attracted much interest in condensed matter physics. In these TI materials, the bulk has a full energy gap while the surface exhibits an odd number of Dirac-cone electronic states, where the spin of the surface electrons is uniquely locked to their momentum \cite{RMP, Zhang_RMP}. Therefore, at any given surface of a TI, the surface electrons moving to one direction (e.g. $+k$) will have only spin up electrons available whereas those of moving to $-k$ only have spin down available. This is in contrast to the case in an ordinary superconductor, where at any given direction the conduction electrons will have both spin up and spin down available for the Cooper pairing. Such distinction can give rise to a wide range of exotic physics. Recently, a number of theories have highlighted these possibilities from both the fundamental physics and application point of view \cite{SUSY, Chubukov, Kane_Proximity, Zhang_TSC, Patrick Lee, Sarma, Hosur}. For example, supersymmetry and Majorana fermions are both very interesting physics phenomena predicted in high energy theories that remain unobserved in particle physics experiments. And it has been theoretically predicted, very recently, that such new physics can be realized in a condensed matter setting \cite{SUSY, Kane_Proximity}, if superconductivity can be induced in a spin-helical gas. Moreover, a low energy realization of these phenomena can also be utilized to build the topological qubit for a topological quantum computer, which therefore is also of value in device applications. The first step towards the realization of any of the fascinating theoretical proposals requires a clear demonstration of the helical-Cooper pairing. Helical-Cooper pairing is defined as the superconducting Bose condensation of a spin-momentum locked Dirac electron gas, independent of the bosonic character of the pairing glue \cite{RMP, Zhang_RMP}. To date, a direct experimental demonstration of the helical-Cooper pairing and their magnetic response remain elusive.


In this paper, we use spin- and momentum-resolved photoemission spectroscopy with sufficiently high resolution and at sufficiently low temperature to allow direct evidence for the \textit{helical} Cooper pairing in a \textit{spin-momentum locked} Dirac electron gas. We achieve this through the observation of \textit{momentum}-resolved Bogoliubov quasi-particle spectrum of a topological insulator (Bi$_2$Se$_3$) in proximity to a superconducting NbSe$_2$ substrate. We further systematically investigate the dependence of the helical Cooper pairing in the Dirac electrons upon varying the TI film thickness or doping magnetic impurities. Our data show that the helical superconductivity in the Dirac surface states can be suppressed by introducing time-reversal symmetry breaking magnetic elements. Our observation of helical-Cooper pairing and superconductivity in spin-Dirac electronic gas serves as an important platform for realizing many exotic physics including emergent supersymmetry \cite{SUSY} physics. We also demonstrate a systematic methodology using the combination of spin- and momentum-resolved ARPES and interface transport that can be more generally applied to discover, isolate, and systematically optimize exotic superconductivity in engineered materials. Previous studies of superconductivity in topological insulator settings have been limited to transport and STM \cite{Hor, Mason, Morpurgo, Nitin, Kanigel, LuLi1,Gordon, Leo, Ando, Dong_PRL}

\bigskip
\bigskip
\textbf{Sample layout for Dirac superconductivity studies}

High quality Bi$_2$Se$_3$ / 2H-NbSe$_2$ interface-heterostructures (Figs. 1\textbf{a,b}) are prepared using molecular beam epitaxy growth (MBE). The growth conditions are systematically optimized to enhance the superconductivity signals in our ARPES measurements. In order to protect the Bi$_2$Se$_3$ surface from exposure to atmosphere, an amorphous selenium layer is deposited on top of the TI surface. This layer can be removed \textit{in situ} in our angle-resolved photoemission spectroscopy (ARPES) experiments by annealing the samples. Fig. 1\textbf{c} shows the momentum-integrated ARPES intensity curves over a wide energy window (core-level spectra) taken on a representative 3 quintuple layer (QL) film before and after removing the amorphous selenium capping layer (decapping). High-resolution ARPES measurements on the Bi$_2$Se$_3$ surface are then performed (Fig. 1\textbf{e}). A sharp spectrum for the Dirac surface states is clearly observed, indicating a good surface/interface quality of our heterostructure. Consistent with previous studies of ultrathin TI films \cite{Xue Nature physics QL,Hedgehog, QL}, we observe a gap at the Dirac point because of the hybridization between the top and bottom surfaces. Furthermore, we perform spin-resolved ARPES measurements (photon energy 50 eV) on the 4QL sample (Fig. 1\textbf{f}). Our spin-resolved measurements confirm that the surface states are indeed singly degenerate near the Fermi level, which is at an energy level far away from the hybridization gap ($v{\cdot}k_{\textrm{F}}>\Delta_{\textrm{hybr}}$) \cite{Hedgehog, QL}. At the Fermi level, a left-handed spin-momentum locking profile is observed, which is one of the critical ingredients for the helical-Cooper pairing as we will show in later sections. 

\bigskip
\bigskip
\textbf{Temperature dependent pairing gap in helical surface states}

In order to study the possible proximity induced superconductivity in the Dirac surface states, we perform systematic ultra-low temperature ($T\sim1$ K) and ultra-high energy resolution ($\sim2$ meV) ARPES measurements on these TI/superconductor heterostructures. We start with the 4QL sample using incident photon energy of 18 eV. Fig. 2\textbf{b} shows the measured dispersion of the Bi$_2$Se$_3$ film. Both the topological surface states (TSSs) and the bulk conduction bands are observed. Six representative momenta, namely ${\pm}{\hbar}k_1$, ${\pm}{\hbar}k_2$, and ${\pm}{\hbar}k_3$ are chosen for detailed studies, where $k_1=0.12$ $\textrm{\AA}^{-1}$ corresponds to the TSSs, and $k_2=0.08$ $\textrm{\AA}^{-1}$ and $k_3=0.04$ $\textrm{\AA}^{-1}$ correspond to the outer and inner parts of the bulk band states, respectively. In order to search for possible superconductivity signals, we study the ARPES energy-spectra at various momentum space locations in close vicinity to the Fermi level ($E_{\textrm{B}}=E_{\textrm{F}}\pm5$ meV). Fig. 2\textbf{c} shows the ARPES spectra at the momentum of $k_1$ (TSSs) at different temperatures. Clear leading-edge shifts (superconducting gap) and coherence peaks are observed at low temperatures. The observed superconducting signals as temperature increases disappear at higher temperatures such as $T=7$ K and 12 K. In order to better visualize the superconductivity gap in our data, the ARPES spectra are symmetrized with respect to the Fermi level, where the temperature evolution of the full (symmetrized) superconducting gap and the coherence peaks are clearly seen in Fig. 2\textbf{d}. These measurements show the existence of induced superconductivity in the helical Dirac electrons occurring in the Bi$_2$Se$_3$ TSSs, which is not possible in conventional momentum-integrated experiments that lacking spin resolution \cite{Hor, Mason, Morpurgo, Nitin, Kanigel, Dong_PRL, LuLi1, Gordon, Leo, Ando}. We compare and contrast the proximity induced superconductivity in the Dirac surface states to that of the bulk band states. Fig. 2\textbf{e} shows the ARPES spectra at $k_2$, where the bulk conduction bands are identified. Superconducting signals including leading edge shifts and coherence peaks are also observed at $k_2$. In order to obtain the magnitude of the superconducting energy gap, we fit the bulk state (${\pm}{\hbar}k_2$, and ${\pm}{\hbar}k_3$) data by the Dynes function \cite{BCS} (black curves in Fig. 2\textbf{g}), which is widely used in $s$-wave superconductors, whereas the surface state data (${\pm}{\hbar}k_1$) is fitted by a BCS function with consideration of the spin-momentum locking and Dirac dispersion properties of the TSSs (blue curves in Fig. 2\textbf{g}). 

Since the surface states and the bulk conduction bands co-exist at the chemical potential (Fig. 2\textbf{b}), we need to examine whether the observed superconducting proximity signal at $k_1$ has contribution from the bulk bands. To further isolate the signals of Dirac surface states from the bulk bands, we choose another incident photon energy of 50 eV where we utilize the photoemission matrix element effect to suppress the spectral weight of the bulk conduction states. As shown in Fig. 2\textbf{i}, at photon energy of 50 eV, the bulk conduction band is almost completely suppressed and the only dispersive band near the Fermi level is the Dirac surface state. We subsequently study the spectra at the momentum $k_1$ where the ARPES signal is dominated by the contribution from the surface states. Leading edge shifts and coherence peaks are clearly observed from $k_1$, which confirms the superconductivity in the Dirac surface states using a different photon energy. These systematic momentum-resolved measurements clearly show the existence of a superconducting helical electron gas, which is realized on the top surface of Bi$_2$Se$_3$ grown on top of an $s$-wave superconductor NbSe$_2$.

We perform ARPES measurements around the surface state Fermi surface as a function of Fermi surface azimuthal angle $\theta$, to study the extent of anisotropy of the surface state superconducting gap. To isolate the surface state signal from the bulk, photon energy of 50 eV is used, where only surface states are observed near the Fermi level as seen in Fig. 3\textbf{a}. Five representative momentum space cut-directions ($\theta1-\theta5$) are chosen as indicated by the dotted lines in Fig. 3\textbf{a}. The helical-surface state superconducting gap observed by ARPES at different $\theta$ angles and their fits are shown in Fig. 3\textbf{b}. The reasonably good surface state fitting results (blue curves in Figs. 2\textbf{g} and 3\textbf{b}) indicate that the obtained surface state superconductivity is consistent with its spin-helical and linear dispersive properties, which supports its helical-Cooper pairing nature. The obtained magnitude of superconducting order parameter is then plotted as a function of Fermi surface angle in Fig. 3\textbf{c}. The superconducting gap is found to be nearly isotropic, which is also consistent with the time-reversal invariant helical nature of the surface state superconductivity as expected theoretically \cite{Kane_Proximity, Zhang_TSC}. We note that the helical Cooper pairing in the topological surface states as observed here is also different from that of in other singly degenerate (spin-momentum locked) but non-topological systems. For example, in a Rashba-2DEG, although the electrons are also singly degenerate, but along any $k-$direction, there are still both spin up and spin down electrons available for Cooper pairing (see Fig. 3\textbf{e}). Therefore, only in the topological surface state, there are electrons with only one spin available for the Cooper pairing along any $k-$direction (Fig. 3\textbf{d}). This unique property as demonstrated here (which we refer as the helical Cooper pairing) is indispensable for all the fascinating phenomena as predicted in theories \cite{SUSY, Chubukov, Kane_Proximity, Zhang_TSC, Patrick Lee, Sarma, Hosur}. In Fig. 3\textbf{f}, we present model calculation analysis, which shows a $p_x{\pm}ip_y$ superconducting order parameter in the Dirac surface states, supporting the helical Cooper pairing nature. Therefore, through our systematic ARPES measurements with simultaneous energy, momentum and spin resolution, we observe superconductivity in an odd number of spin-momentum locked Dirac surface states, which serves as the direct experimental evidence for the helical-Cooper pairing.

\bigskip
\bigskip
\textbf{Dependence on surface-surface hybridization}

We study the observed superconductivity in the surface states as a function of surface-to-surface hybridization strength (effectively as a function of the TI film thickness), in order to experimentally prove its proximity-induced nature. As shown in Fig. 4\textbf{a}, a sample with a Dirac point hybridization gap as large as $\sim200$ meV is realized in a 3QL film sample. Clear leading-edge shifts, coherence peaks and their temperature evolution are observed in the TSSs near the Fermi energy (Figs. 4\textbf{b,c}), evidence for the helical-Cooper pairing and helical superconductivity in the 3QL film samples. The observed surface state superconducting gap value is about $0.7-0.8$ meV. We now turn to a gapless (7QL) sample as shown in Fig. 4\textbf{d}. The absence of hybridization gap at the Dirac point reveals that surface state wavefunctions from the top and the interface surfaces are completely separated in real space. Finite superconductivity signals are observed in the ARPES spectra at both momenta of $k_1$ (Dirac surface states) and $k_2$ (bulk conduction states), which are found to be weaker than the gapped samples. The Bi$_2$Se$_3$ top surface's superconducting gap as a function of Dirac point gap value (surface-to-surface hybridization strength) is shown in Figs. 4\textbf{g,h} in a (surface and bulk) band-resolved fashion.  It can be seen that the induced superconducting gap near the top surface increases with a larger the Dirac point hybridization gap, which is realized in thinner TI films.  This observation is qualitatively consistent with the theoretical description of the superconducting proximity effect, where the Cooper pair potential on the top surface is enhanced with the decreasing thickness of the normal metal. More interestingly, it can be seen that the surface state superconducting gap increases at a faster rate than that of the proximity gap on the bulk band. Such contrast reveals that stronger surface-to-surface hybridization significantly enhances the helical pairing in the surface states on the top surface. These microscopics of the superconducting proximity effect observed in our data will be a valuable guide in properly interpreting the vast complexity of the transport data addressing the proximity effects in TI films and heterostructures.

\bigskip
\bigskip
\textbf{Effects of time-reversal symmetry breaking on helical superconductivity}

In order to test the time-reversal invariant character of surface state superconductivity of Bi$_2$Se$_3$ required by its helical nature, we study manganese (Mn) doped Bi$_2$Se$_3$ grown on top of NbSe$_2$. Mn atoms are introduced into Bi$_2$Se$_3$ throughout the film during the MBE growth. ARPES studies on 4QL Bi$_2$Se$_3$ films with two different Mn doping (4\% and 10\%) levels are presented in Figs. 5\textbf{a-f}. Bulk manganese doping is found to hole dope the system, thus bringing the chemical potential closer to the Dirac point. In principle, Mn impurities can affect the induced superconductivity in several different channels, all leading to the suppression of proximity superconductivity: The major effect is that Mn impurities introduce (either random or ordered) magnetism into the TI, which is destructive to the helical pairing; Another minor effect, which can also contribute, is that impurities generate random disorder reducing the electron mean free path. As shown in Figs. 5\textbf{a-f}, the superconducting coherence peak is strongly suppressed in the heavily Mn-doped samples. These momentum-resolved measurements allow us to isolate the strength of the effect on helical surface states and bulk bands, thus directly demonstrating that magnetic impurities lead to strong pairing breaking in both conventional and helical pairing channels. The complete suppression of superconductivity in the helical Dirac surface states upon strong Mn doping effectively drives a topological phase transition from a helical superconductor to a normal Dirac metal state as seen in our data. A sample fabricated to lie near the critical point of this transition can host many exotic phenomena (which will be discussed later in this paper).

\bigskip
\bigskip
\textbf{Point-contact Andreev-reflection transport}

In order to check that the top surface superconductivity is indeed a proximity effect, we use the well-established methodology  of point-contact Andreev reflection measurements to compare the induced gap values at the NbSe$_2$/Bi$_2$Se$_3$ heterointerface. The point-contact spectra on a NbSe$_2$/Bi$_2$Se$_3$ (16QL)  heterostructure sample are shown in Fig. 6\textbf{a}. At temperatures below the $T_{\textrm{c}}$ ($\sim7.2$ K) of NbSe$_2$, the differential conductance ($dI/dV$) around zero bias increases as a result of the Andreev reflection process, similar to the bare NbSe$_2$ spectra. Interestingly, a second differential conductance increase appears below $\sim5$ K. The sharp rise of differential conductance corresponds to the energy gap of the superconducting layer. From the data, we obtain the larger gap ($\Delta_1$) changing from 0 to $1.3\pm0.2$ mV from 7.5 K to 3.0 K.  The second gap ($\Delta_2$) feature changes from 0 to $0.8\pm0.2$ mV from 5.0 K to 3.0 K. Similar two gap features have been observed in point-contact studies of Ag/Pb \cite{Ag_Pb} and Si/Nb \cite{Si_Nb} interface samples and were attributed to the superconducting energy gap of the superconducting layer and the proximity induced gap in the normal metal layer at the N/S interface, respectively. Injected electrons from the point-contact are Andreev reflected inside the superconducting Bi$_2$Se$_3$ proximity layer if their energies are lower than the induced superconducting gap in Bi$_2$Se$_3$. When their energies are above the induced gap $\Delta_{\textrm{TI}}$ but below the NbSe$_2$ superconducting gap $\Delta_{\textrm{SC}}$, injected electrons are not affected by the order parameter in the Bi$_2$Se$_3$ layer but Andreev reflected in the NbSe$_2$ region. Therefore, the edge around $\Delta_{1}$ is likely due to the NbSe$_2$ gap, while the sharp edge around $\Delta_2$ in the conductance spectrum is likely to reflect the induced gap in Bi$_2$Se$_3$ near the interface. It is worth noting that unlike ARPES, which is mostly sensitive to the top surface, the point-contact transport probes deeper into the superconductor, similar to the electron mean free path, which is estimated to be $\sim$16 nm in our films. The induced gap in Bi$_2$Se$_3$ at the Bi$_2$Se$_3$/NbSe$_2$ interface is $\sim$0.8 meV at 3 K from the point-contact measurement, which is in reasonable agreement with the fitted gap value extracted from the ARPES measurement (see data on 3QL sample). Our results thus suggest that the combination of ARPES and point-contact transport together provides a powerful method for probing superconducting proximity effect which can be used to correlate the proximity gap on the top surface and the buried interface if film thickness is not too large (not larger than the superconducting coherence length).

\bigskip
\bigskip

\section{Discussions}

In contrast to idealized theoretical models \cite{Kane_Proximity, Zhang_TSC} of topological superconductivity where only Dirac surface states cross the Fermi level, real samples exhibit a complex phenomenology due to the coexistence of multiple bands at the chemical potential, as demonstrated in our data above. Thus, the interpretation of  experimental studies must take into account both the desirable Cooper pairing from the Dirac surface states and conventional superconductivity from the bulk, trivial surface states and impurity surface states. The coexistence of multiple bands at the Fermi level means that any superconductivity realized in actual TI materials consist of not only the desirable helical Cooper pairing from the Dirac surface states but also conventional superconductivity from the bulk states, as shown above in our data. We note that although progress has been reported by using conventional transport and STM experiments \cite{Hor, Mason, Morpurgo, Nitin, Kanigel, LuLi1,Gordon, Leo, Ando, Dong_PRL}, those studies do not have the spin and momentum resolution necessary to distinguish the helical Cooper pairing from that of the conventional superconductivity from other bands that intermix at the interface making interpretation of Majorana fermions unreliable or complex. Hence a direct experimental demonstration of the existence of superconductivity in the helical Cooper pairing channel remained elusive before our momentum and spin space observations reported here. In fact, it has been recently shown both theoretically and experimentally \cite{Patrick Lee2, TeWari, Marcus, Franceschi} that the conventional superconductivity in the bulk and impurity bands at the interface or surface lead to ambiguous interpretations of the transport and STM data. In order to achieve a clear case for Majorana zero mode, the helical component of the Cooper pairing must be isolated, as demonstrated here. Therefore, it is in this context that our observation of helical Cooper pairing and superconductivity in a half Dirac gas is of critical importance. Additionally, the overall methodology employed here can be applied to isolate helical Cooper pairing in other systems and in connection to a feedback loop for material growth for the optimization of the helical channel.


We also note that our systematic studies (by observing the superconducting gap (leading edge shift), the clear coherence peak, as well as their systematic dependence upon varying temperature, TI film thickness and doping magnetic impurities) are in contrast to the debatable ARPES results on Bi$_2$Se$_3$/BSCCO samples \cite{Zhou, BSCCO_Hasan, BSCCO_Valla}. In that case, no superconducting coherence peak was observed \cite{Zhou, BSCCO_Hasan, BSCCO_Valla}, and the claim of a $\geq15$ meV leading edge shift \cite{Zhou} in the Dirac surface states in Bi$_2$Se$_3$/BSCCO is in contrast to the absence of any observable leading edge shift in the other two studies \cite{BSCCO_Hasan, BSCCO_Valla}. In fact, a strong superconducting proximity effect is inconsistent with important facts including the severe mismatch of both Fermi momenta and crystal symmetries between Bi$_2$Se$_3$ and BSCCO, very short out-of-plane superconducting coherence length of high-$T_{\textrm{c}}$ superconductors, as well as the different superconducting pairing symmetries between a TI and a $d-$wave cuprate superconductor. Thus, our data strongly supports the view that TI/NbSe$_2$ is a more ideal platform than TI/BSCCO for the proposed novel physics if the system can be further optimized increasing helical Cooper pairing channel by tuning the material parameters.

We discuss the emergent topological phenomena that can be enabled by our identification of helical-Cooper pairing. One exciting scenario is to realize supersymmetric phenomenon in our experimental setup \cite{SUSY} by further improvement of the film quality and magnetic doping process. As shown in Fig. 6\textbf{d}, magnetic doping or an external in-plane magnetic field is necessary to drive the system to the critical point between the helical superconductivity and the normal Dirac gas states, with the chemical potential tuned to the Dirac point (demonstrated in Fig. 6\textbf{b}). Under this condition, theory predicts that topological surface states (a fermionic excitation) and the fluctuations of superconducting order (a bosonic excitation) satisfy supersymmetry relationship, and therefore, strikingly, possess the same Fermi/Dirac velocity and same lifetime or self-energy \cite{SUSY}. While the superpartners of elementary particles in high energy physics have never been experimentally observed, the experimental methodologies, artificial sample fabrication control and experimental observations reported here pave the way for simulating and testing supersymmetric physics concepts in future sub-Kelvin nanodevices fabricated out of sample configurations discussed here.

\bigskip
\bigskip

\section{Methods}
Single crystalline 2H-NbSe$_2$ samples ($\sim3$ mm $\times$ $3$ mm) were grown using the iodine-vapour-transport method \cite{NbSe2}, where a high residual resistivity ratio $(\textrm{RRR}=\frac{\rho_{300\textrm{K}}}{\rho_{0\textrm{K}}}\sim100)$ was achieved. The NbSe$_2$ crystals were cleaved \textit{in situ} under ultra-high vacuum, and high quality topological insulator Bi$_2$Se$_3$ thin films were then grown on top of freshly cleaved surface of NbSe$_2$ crystals using standard MBE method. 

Ultra-low temperature and ultra-high energy resolution ARPES measurements were performed using the UE$112$-lowE-PGM-b$+1^3$ ARPES beamline in the BESSY II storage ring in Berlin, Germany, which is equipped with a VG Scienta 4000 electron analyzer and $^3$He cryo-manipulator, using incident photon energy range from 17 eV to 80 eV, with the lowest achievable sample temperature of about 0.9 K, and the base pressure of $<5\times10^{-11}$ torr. The total energy and momentum resolution were set to $\sim2.4$ meV and $0.01$ $\textrm{\AA}^{-1}$. Spin-resolved ARPES measurements were performed at the PHOENEXS chamber at the UE112-PGM1 beamline at BessyII. These measurements were performed with the linearly $p$-polarized light at synchrotron radiation photon energies from 50 eV to 70 eV, where the final state effects are demonstrated to be negligible \cite{Rader}. Additional ARPES measurements characterizing the overall electronic structure of Bi$_2$Se$_3$/NbSe$_2$ (including the NO$_2$ adsorption experiments) were also performed with $29$ eV to $64$ eV incident photon energies at the beamlines 4.0.3, 10.0.1 and 12.0.1 at the Advanced Light Source (ALS) in the Lawrence Berkeley National Laboratory (LBNL) in Berkeley, CA. The base temperature and base pressure of the ARPES beamlines at the ALS were about 15 K and $<5\times10^{-11}$ torr, and the total energy and momentum resolution of these beamlines were about 15 meV and $0.01$ $\textrm{\AA}^{-1}$.

\section{Acknowledgement}

The work at Princeton and Princeton-led synchrotron X-ray-based measurements is supported by U.S. DOE DE-FG-02-05ER46200. MBE growth of Bi$_2$Se$_3$/NbSe$_2$ samples at Penn State was supported by ARO through the MURI porogram. The point-contact measurements were supported by U.S. DOE DE- FG02-08ER4653. C.F and M.J.G. are supported by the ONR under grant N00014-11-1-0728 and N00014-14-1-0123 and B.D. and M.J.G. are supported by the AFOSR under grant FA9550-10-1-0459. FCC acknowledges the support provided by MOST-Taiwan under project number NSC-102-2119-M-002-004

\newpage

\noindent
\textbf{\large{Figure Captions}}

\bigskip

\noindent
\textbf{Figure 1 $|$ Topological superconductivity via the proximity effect}. \textbf{a,} A schematic layout of ultra-thin Bi$_2$Se$_3$ films epitaxially grown on the (0001) surface of single crystalline $s$-wave superconductor 2H-NbSe$_2$ ($T_{\textrm{c}}=7.2$ K) using the MBE technique. \textbf{b,} High resolution transmission electron microscopy (TEM) measurements of the Bi$_2$Se$_3$/NbSe$_2$ interface at 200 keV electron energy. An atomically abrupt transition from NbSe$_2$ layered structure to the layered quintuple layer structure of Bi$_2$Se$_3$ is resolved, showing a good atomically flat interface crystal quality. \textbf{c,} Momentum-integrated ARPES intensity curves over a wide binding energy window (core-level spectra) taken on a representative 3QL Bi$_2$Se$_3$ ($\simeq3$ nm) film grown on NbSe$_2$ before and after removing the amorphous selenium capping layer (decapping). \textbf{d,} A low-energy electron diffraction (LEED) image on a 4QL Bi$_2$Se$_3$ film shows six-fold pattern providing evidence that the thin Bi$_2$Se$_3$ film is well-ordered. \textbf{e,} High-resolution ARPES dispersion map of a 4QL Bi$_2$Se$_3$ film on NbSe$_2$ after decapping using incident photon energy of 50 eV. The white circle and cross schematically show the measured direction of the spin texture on the top surface of our 4QL Bi$_2$Se$_3$ film shown in Panel (\textbf{f}). \textbf{f,} Spin-resolved ARPES measurements on 4QL Bi$_2$Se$_3$ as a function of binding energy at a fix momentum which is indicated by the white dotted line in Panel (\textbf{e}). \textbf{g,} High-resolution ARPES dispersion map of a 6QL Bi$_2$Se$_3$ film on NbSe$_2$ at $T=12$ K. The white arrow indicates the momentum for the temperature dependent EDC in Panel (\textbf{h}). \textbf{h,} Temperature dependence of the ARPES spectra in Panel (\textbf{g}).
\bigskip
\bigskip

\noindent
\textbf{Figure 2 $|$ Spectroscopically-resolved proximity-induced 2D topological superconductivity.} \textbf{a,} ARPES dispersion maps of Bi$_2$Se$_3$/NbSe$_2$ as a function of Bi$_2$Se$_3$ film thickness. All dispersion maps are measured with photon energy of 50 eV, except the 7QL sample which is measured by 18 eV. The blue arrows quantitatively depict the spin texture configuration in the ultra-thin limit. The length of the arrow is proportional to the magnitude of the spin polarization. \textbf{b,} ARPES dispersion map of a 4QL Bi$_2$Se$_3$ film measured at $T=12$ K using incident photon energy of 18 eV. \textbf{c,} ARPES spectra at the fixed momentum of $k_1$ (the topological surface states). \textbf{d,} Symmetrized ARPES spectra at $k_1$. \textbf{e,} ARPES spectra at the fixed momentum of $k_2$ (bulk band states). \textbf{f, g,} Symmetrized ARPES spectra at ${\pm}k_1$, ${\pm}k_2$, and ${\pm}k_3$ at $T\sim1$ K. The surface state gap (${\pm}k_1$) is fitted by a BCS function considering its spin-momentum locking and Dirac dispersion properties, where as the bulk gap is fitted by the Dynes function \cite{BCS}. \textbf{h,} ARPES spectrum of the \textit{in situ} evaporated gold film, where the kinetic energy of the Fermi level and the energy resolution are determined. \textbf{i,} ARPES dispersion map of a 4QL Bi$_2$Se$_3$ film measured with incident photon energy of 50 eV. \textbf{j,k,} ARPES spectra at the fixed momentum of $k_1$. The insets of Panels (\textbf{c, j}) show an ARPES dispersion map near the Fermi level at low temperature of $T\sim1$ K at photon energies of 18 eV and 50 eV, respectively.

\bigskip
\bigskip

\noindent
\textbf{Figure 3 $|$ Surface superconducting gap and helical pairing.} \textbf{a,} Fermi surface map taken at incident photon energy of 50 eV. The white dotted lines indicate the momentum-space cut-directions chosen to study the surface state superconducting gap as a function of Fermi surface angle $\theta$ around the surface state Fermi surface. \textbf{b,} Symmetrized and normalized ARPES spectra along $\theta_1$ through $\theta_5$ respectively (red) and their surface gap fittings. \textbf{c,} Fermi surface angle dependence of the estimated superconducting gap around the surface state Fermi surface. \textbf{d}, Illustrations for helical-Cooper pairing in a spin-momentum locked helical electron gas and \textbf{e,} The conventional $s$-wave Cooper pairing in an ordinary superconductor. Note that the superconductivity in a Rashba-2DEG can be visualized also in the same schematic (Panel \textbf{e}) but with the length of the $k$-vector for spin up and spin down electrons being different (spin-split). However, for a Rashba-2DEG, it is still true that along any $k$-direction, both spin up and spin down electrons cross the Fermi level and are available for the Cooper pairing. \textbf{f,} Model calculation results of a topological insulator film in proximity to an $s$-wave superconductor shows the calculated total superconducting pairing amplitude and its decomposed singlet ($\widetilde{\Delta}_S$) and triplet ($\widetilde{\Delta}_T$) components on the top surface of a 4-unit-cell thick TI interfaced with an $s$-wave superconductor, which further confirms the helical (topologically nontrivial) nature of the induced surface state superconductivity in our Bi$_2$Se$_3$ films. Since $\widetilde{\Delta}$ is a dimensionless number, we denote a tilde $\widetilde{\Delta}$ to differentiate from the superconducting gap $\Delta$ measured in experiments. 

\bigskip
\bigskip

\noindent
\textbf{Figure 4 $|$ Hybridization dependence of superconducting gap.} \textbf{a,} ARPES dispersion map of a 3QL Bi$_2$Se$_3$ film measured at $T=12$ K using incident photon energy of 50 eV. \textbf{b,} ARPES spectra at the fixed momentum of $k_1$ at different temperatures. \textbf{c,} Symmetrized ARPES spectra at different temperatures. \textbf{d,} ARPES dispersion map of a 7QL Bi$_2$Se$_3$ film measured at $T=12$ K using incident photon energy of 18 eV. \textbf{e, f,} ARPES spectra at the fixed momenta of $k_1$ and $k_2$. \textbf{g,} ARPES measured superconducting gap for topological surface states ($k_1$) and for the bulk conduction states ($k_2$) as a function of Dirac point gap value (surface-to-surface hybridization strength). The dotted lines are guides to the eye. \textbf{h,} Surface-to-surface hybridization dependence in calculation. Since the pairing amplitude $\widetilde{\Delta}_{\textrm{TI}}$ in calculation is a dimensionless number, we normalize it by the pairing amplitude of the substrate superconductor $\widetilde{\Delta}_{\textrm{Substrate}}$ (a constant).

\bigskip
\bigskip

\noindent
\textbf{Figure 5 $|$ Destruction of the helical Cooper pairing via time-reversal symmetry breaking magnetic doping.} \textbf{a,} ARPES dispersion map of a 4QL Mn($4\%$)-doped Bi$_2$Se$_3$ film measured at $T\sim1$ K. \textbf{b, c,} ARPES spectra at the fixed momenta of $k_1$ (topological surface states) and $k_2$ (bulk conduction states). \textbf{d-f,} Same as panels (\textbf{a-c}) but for 10\% Mn doping. The Mn concentrations indicated here are nominal, which means they are estimated by the flux ratio of Mn flux : Bi flux during the MBE growth. 

\bigskip
\bigskip

\noindent
\textbf{Figure 6 $|$ Point-contact interface transport and conditions for theoretically predicted emergent supersymmetry in a 2D topological superconductor.} \textbf{a,} Point-contact transport ($dI/dV$ vs bias voltage) as a function of temperature. For $T=3.5$ K, below the $T_{\textrm{c}}$ of NbSe$_2$, as the bias voltage is swept from $\pm 3$mV to $0$mV, $dI/dV$ first increases, then levels off at $|V|\simeq1.3$ mV, and then increases again, reaching a maximum at $|V|\simeq0.8$ mV. $dI/dV$ for other $T <  T_{\textrm{c}}$ exhibits a similar behavior. This indicates two Andreev reflection channels with different sizes of superconducting gap. The inset illustrates the two Andreev reflection processes via the induced superconducting gap in Bi$_2$Se$_3$ and intrinsic superconducting gap in NbSe$_2$, respectively. \textbf{b,} Measured surface state dispersion upon \textit{in situ} NO$_2$ surface adsorption on the surface of a 7QL Bi$_2$Se$_3$/NbSe$_2$ sample using incident photon energy of 55 eV at temperature of 20 K. The NO$_2$ dosage in the unit of Langmuir ($1\textrm{L}=1\times10^{-6}$ torr${\cdot}$sec) is noted on the top-right corners of the panels, respectively. The white dotted lines in the last panel are guides to the eye. \textbf{c,d,} Theoretically proposed \cite{SUSY} unusual criticality related to supersymmetry phenomena can be realized in the topological insulator/$s$-wave superconductor interface as the chemical potential is tuned to the surface state Dirac point and an in-plane magnetization drives the system to the critical point between superconducting and normal Dirac metal phases. Fermions and bosons are expected to feature the same band velocity and quasi-paticle lifetime \cite{SUSY}, where fermion velocity is estimated to be around $5\times10^5$ m/s from our ARPES result. This cartoon refers to a future possibility.

\clearpage
\begin{figure*}
\centering
\includegraphics[width=18cm]{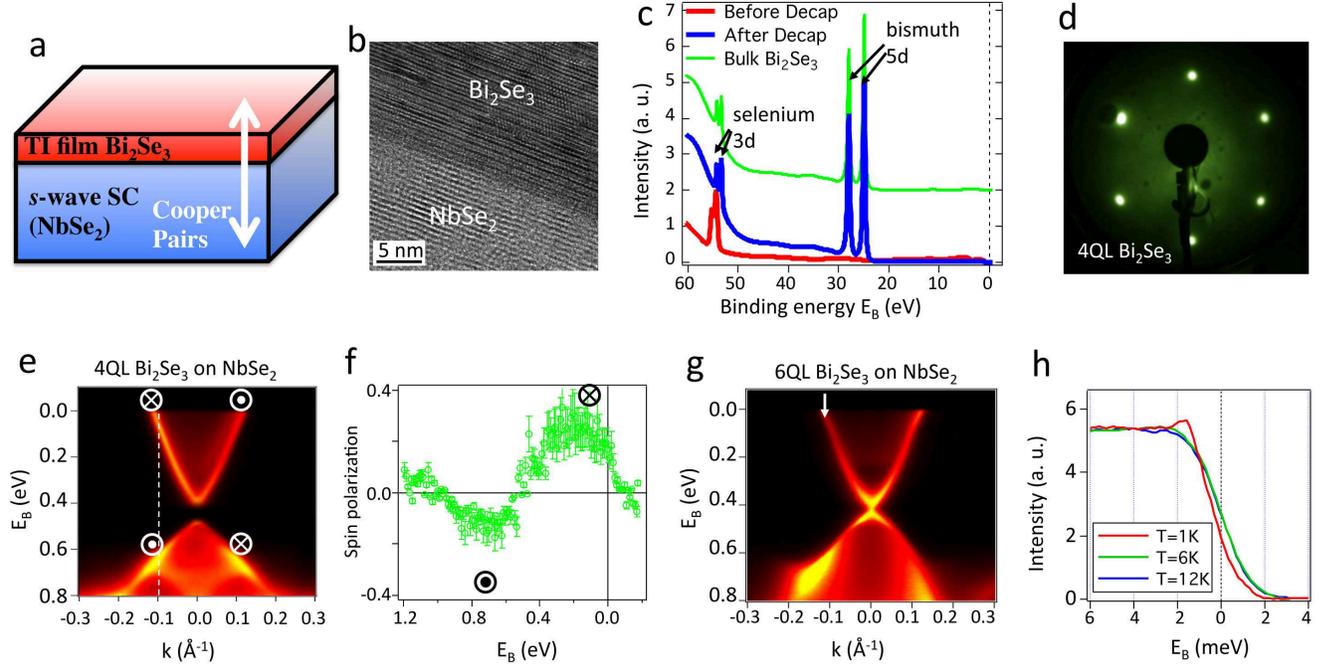}
\caption{\label{Fig1}\textbf{Topological superconductivity via the proximity effect.}}
\end{figure*}

\clearpage
\begin{figure}
\includegraphics[width=18cm]{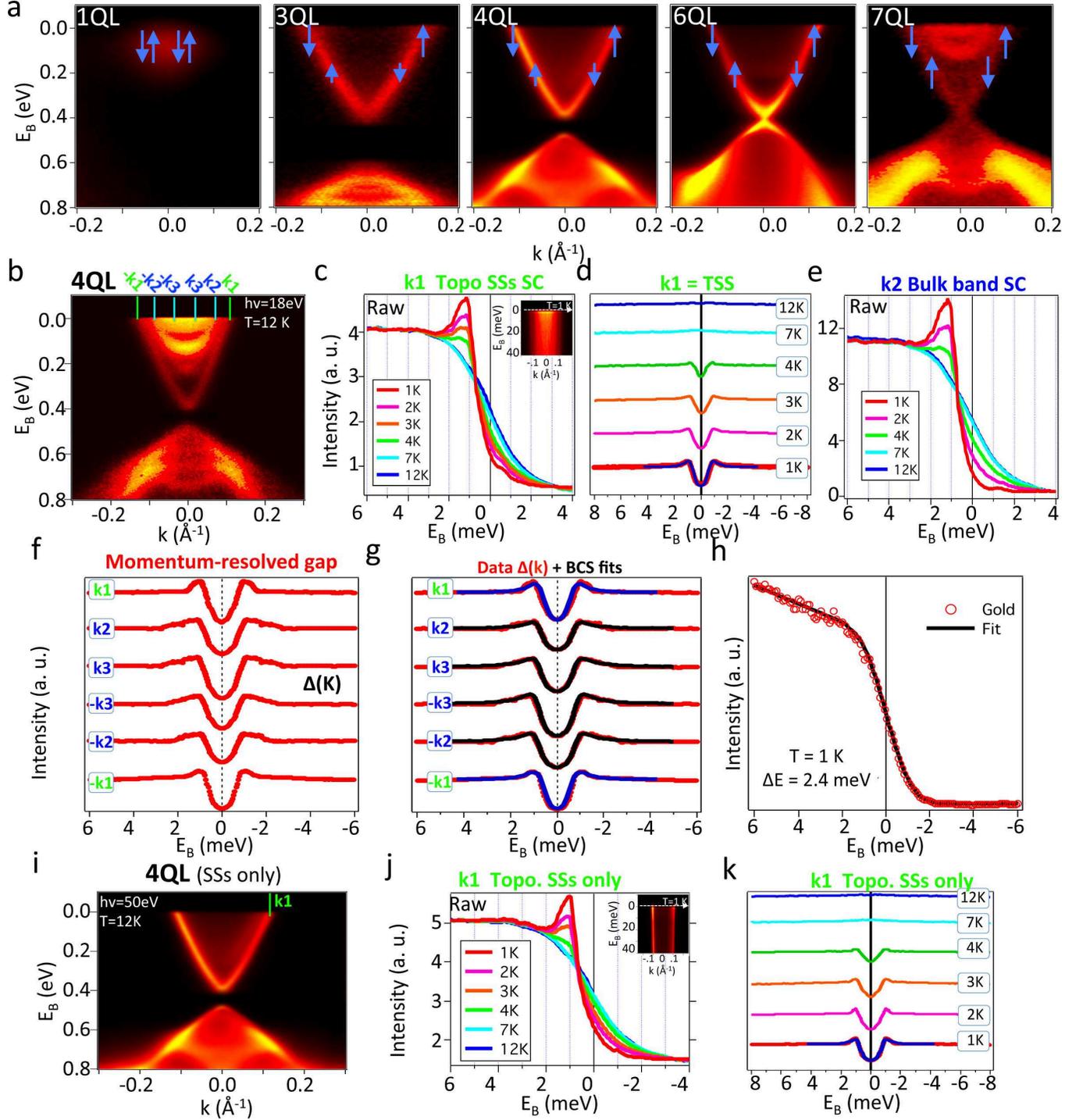}
\caption{\label{Fig2}\textbf{Spectroscopically-resolved proximity-induced 2D topological superconductivity.}}
\end{figure}

\clearpage
\begin{figure}
\centering
\includegraphics[width=18cm]{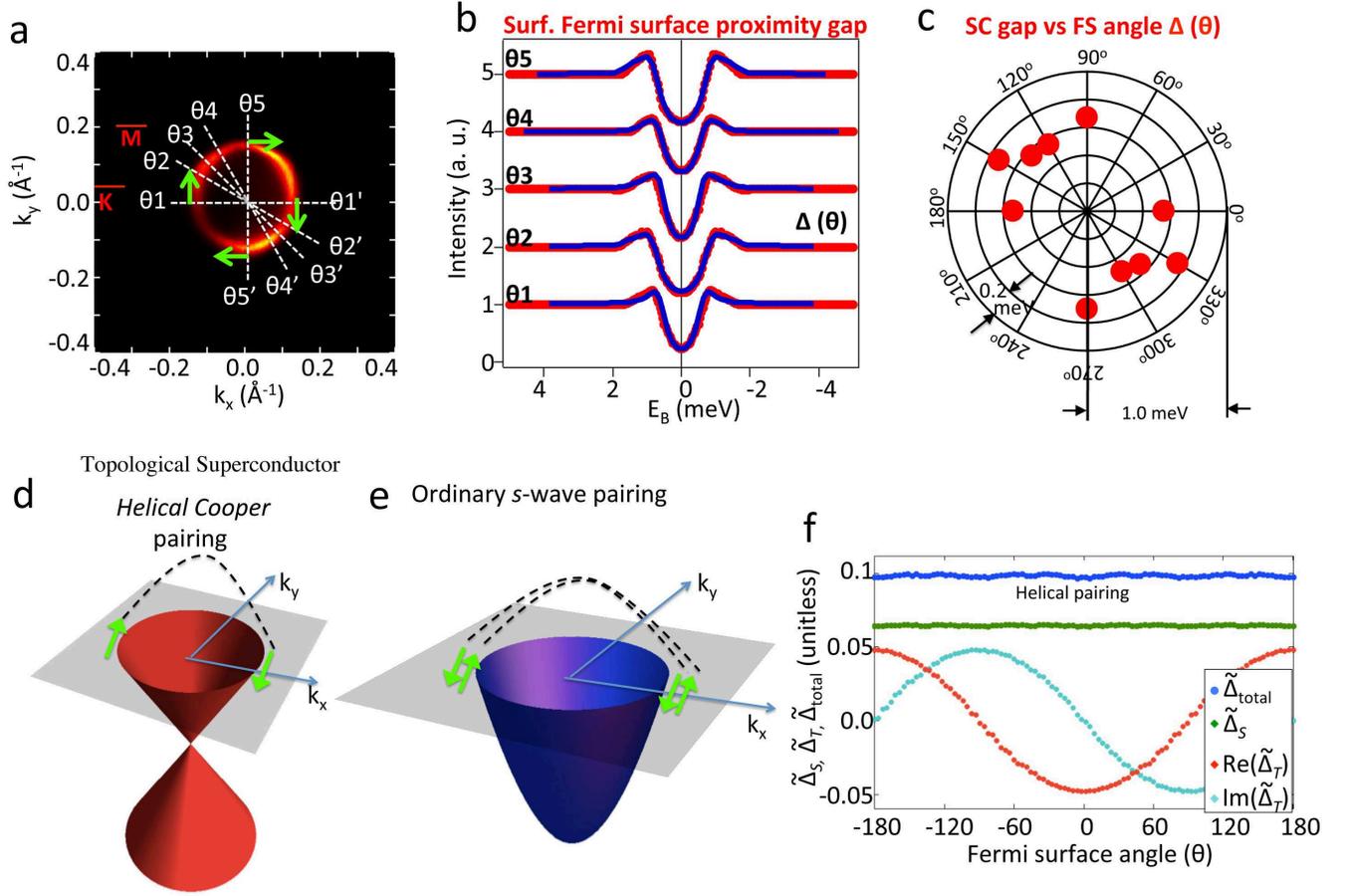}
\caption{\label{Fig3}\textbf{Surface superconducting gap and helical pairing.}}
\end{figure}

\clearpage
\begin{figure}
\centering
\includegraphics[width=18cm]{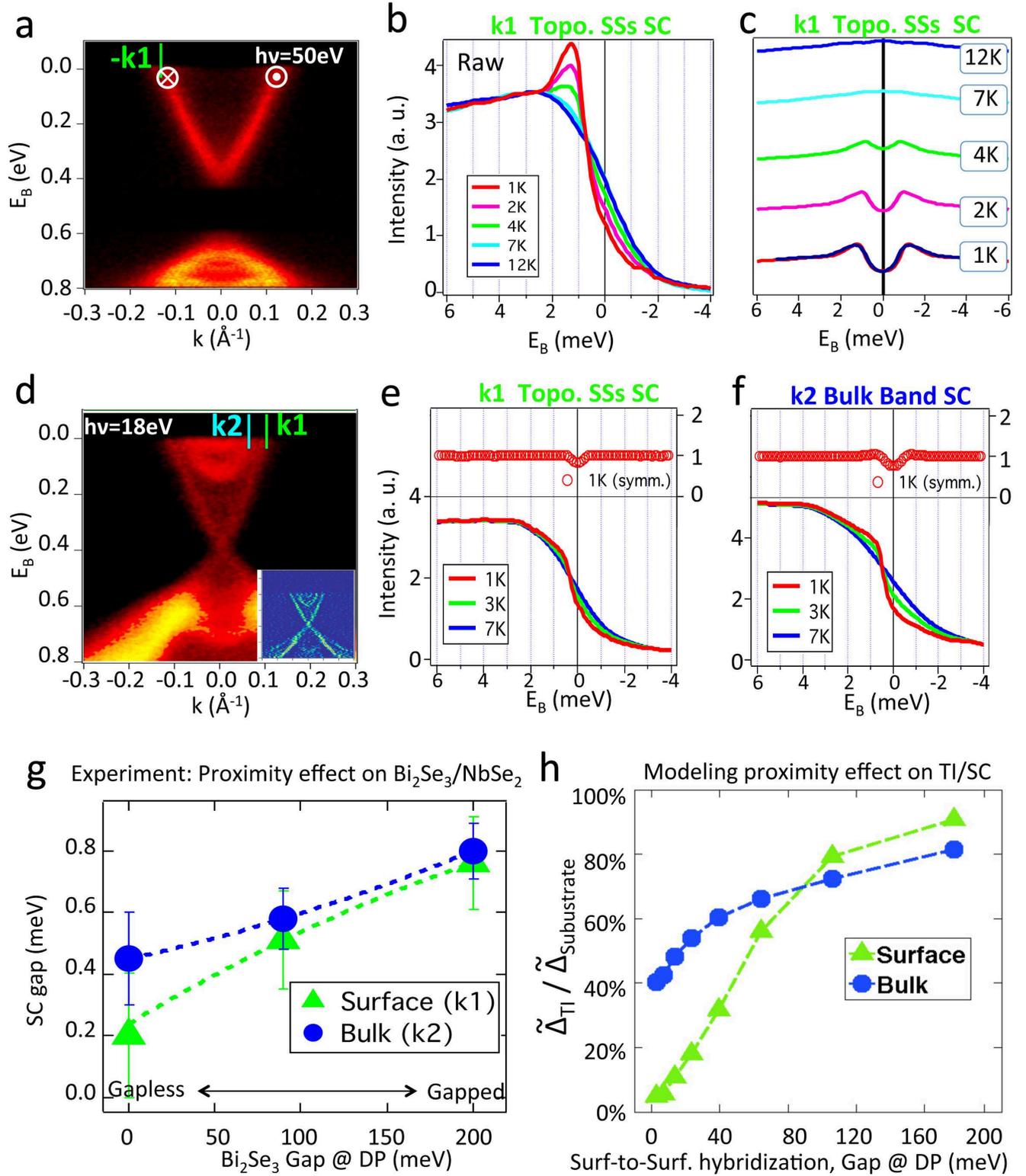}
\caption{\label{Fig4}\textbf{Hybridization dependence of superconducting gap.}}
\end{figure}

\clearpage
\begin{figure}
\centering
\includegraphics[width=18cm]{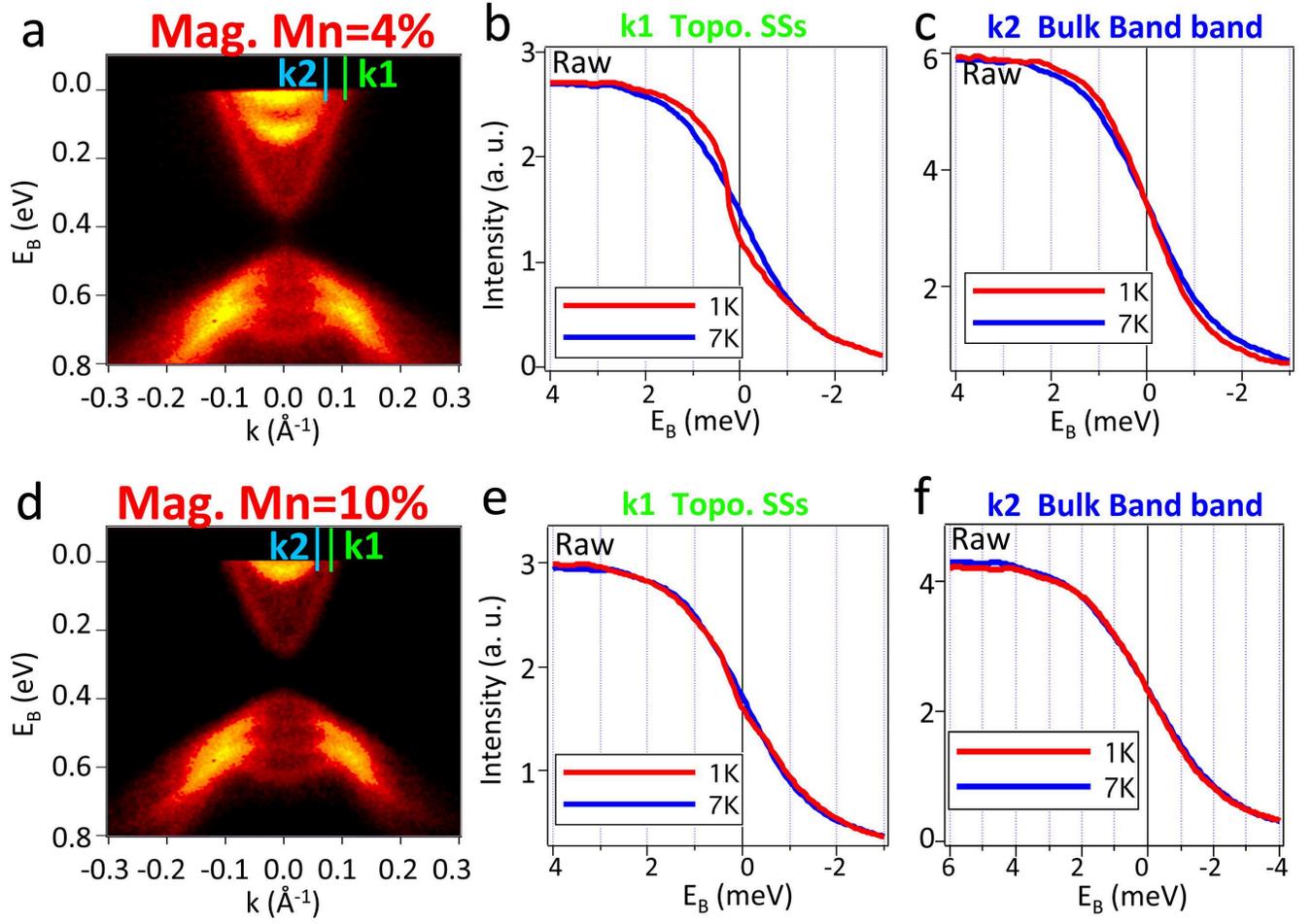}
\caption{\label{Fig5}\textbf{Destruction of the helical Cooper pairing via time-reversal symmetry breaking magnetic doping.}}
\end{figure}

\clearpage
\begin{figure}
\centering
\includegraphics[width=18cm]{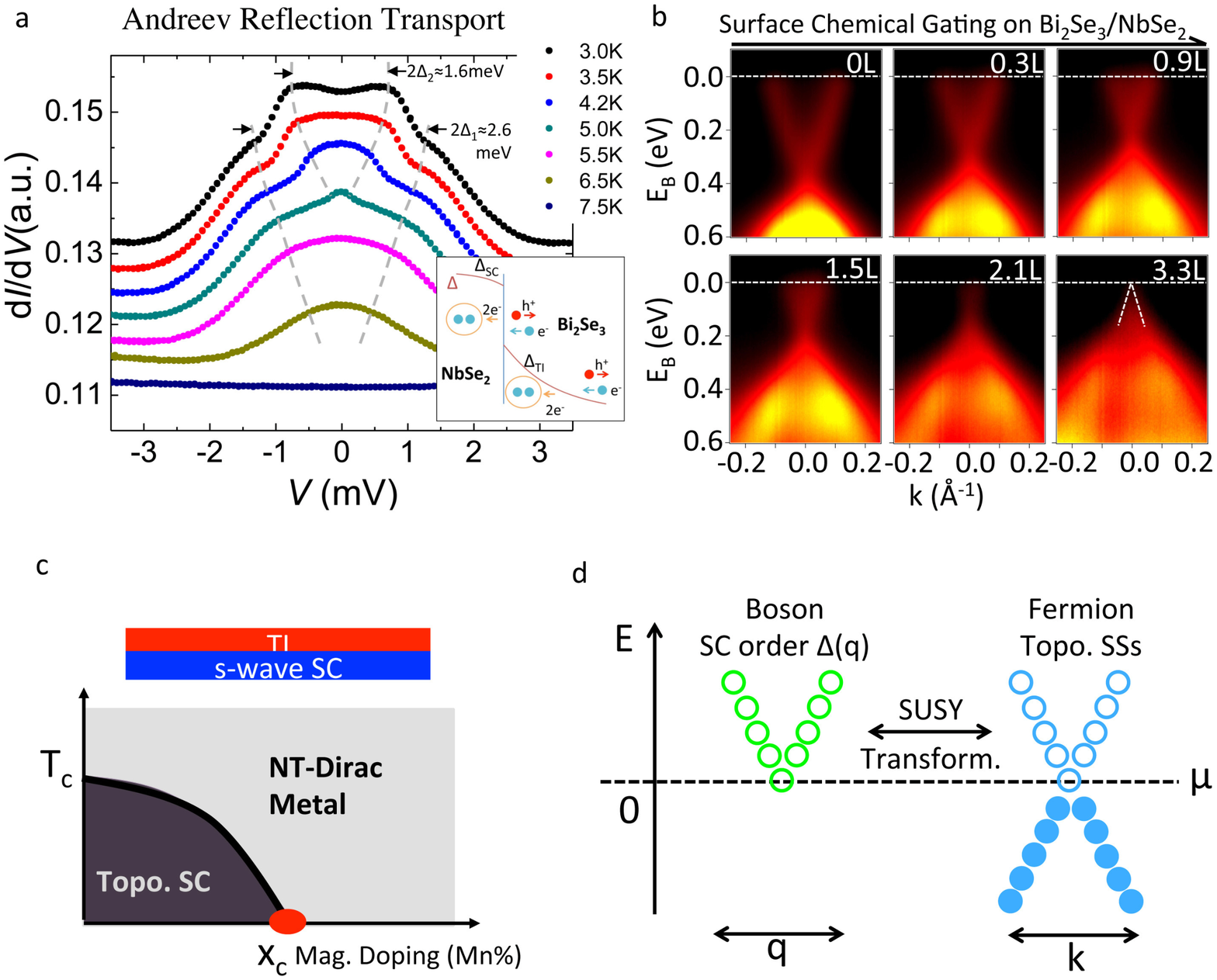}
\caption{\label{Fig6}\textbf{Point-contact interface transport and conditions for theoretically predicted emergent supersymmetry (schematic) in a 2D topological superconductor.}}
\end{figure}


\begin{thebibliography}{21}
\bibitem{TI_book_2014} Hasan, M. Z., Xu, S.-Y. \& Neupane, M. Topological Insulators, Topological Crystalline Insulators, and Topological Kondo Insulators. Preprint at http://arXiv:1406.1040 (2014).

\bibitem{SUSY} Grover, T., Sheng, D. N. \& Vishwanath, A. Emergent Space-Time Supersymmetry at the Boundary of a Topological Phase. \textit{Science} $\mathbf{344}$, 280 (2014).
\bibitem{Kane_Proximity} Fu, L. \& Kane, C. L. Superconducting Proximity Effect and Majorana Fermions at the Surface of a Topological Insulator. \textit{Phys. Rev. Lett.} $\mathbf{100}$, 096407 (2008).

\bibitem{Chubukov} Nandkishore, R., Levitov, L. S. \& Chubukov, A. V. Chiral superconductivity from repulsive interactions in doped graphene. \textit{Nature Phys.} $\mathbf{8}$, 158-163 (2012).



\bibitem{Zhang_TSC} Qi, X.-L., Hughes, T. L., Raghu, S. \& Zhang. S.-C. Time-Reversal-Invariant Topological Superconductors and Superfluids in Two and Three Dimensions. \textit{Phys. Rev. Lett.} $\mathbf{102}$, 187001 (2009).
\bibitem{Patrick Lee} Potter, A. C. \& Lee, P. A. Engineering a $p+ip$ superconductor: Comparison of topological insulator and Rashba spin-orbit-coupled materials. \textit{Phys. Rev. B} $\mathbf{83}$, 184520 (2011).

\bibitem{Sarma} Sau, J. D. \textit{et al}. A generic new platform for topological quantum computation using semiconductor heterostructures. \textit{Phys. Rev. Lett.} $\mathbf{104}$, 040502 (2010).


\bibitem{Hosur} Hosur, P. \textit{et al}. Majorana Modes at the Ends of Superconductor Vortices in Doped Topological Insulators. \textit{Phys. Rev. Lett.} $\mathbf{107}$, 097001 (2011).
\bibitem{RMP} Hasan, M. Z. \& Kane, C. L. Topological insulators. \textit{Rev. Mod. Phys.} $\mathbf{82}$, 3045-3067 (2010).
\bibitem{Zhang_RMP} Qi, X. -L. \& Zhang, S. -C. Topological insulators and superconductors. \textit{Rev. Mod. Phys.} $\mathbf{83}$, 1057-1110 (2011).

\bibitem{Hor} Hor, Y. S. \textit{et al}. Superconductivity in Cu$_x$Bi$_2$Se$_3$ and its implications for pairing in the undoped topological insulator. \textit{Phys. Rev. Lett.} $\mathbf{104}$, 057001 (2010).
\bibitem{Ando} Sasaki, S. \textit{et al}. Topological Superconductivity in Cu$_x$Bi$_2$Se$_3$. \textit{Phys. Rev. Lett.} $\mathbf{107}$, 217001 (2011).
\bibitem{Nitin} Zhang, D. \textit{et al}. Superconducting proximity effect and possible evidence for Pearl vortices in a candidate topological insulator. \textit{Phys. Rev. B} $\mathbf{84}$, 165120 (2011).
\bibitem{Kanigel} Koren, G. \textit{et al}. Proximity-induced superconductivity in topological Bi$_2$Te$_2$Se and Bi$_2$Se$_3$ films: Robust zero-energy bound state possibly due to Majorana fermions. \textit{Phys. Rev. B} $\mathbf{84}$, 224521 (2011).

\bibitem{Morpurgo} Sac\'ep\'e, B. \textit{et al}. Gate-tuned normal and superconducting transport at the surface of a topological insulator. \textit{Nature Comm.} $\mathbf{2}$, 575 (2011).

\bibitem{LuLi1} Qu, F. \textit{et al}. Strong Superconducting Proximity Effect in Pb-Bi$_2$Te$_3$ Hybrid Structures. \textit{Scientific Reports} $\mathbf{2}$, 339 (2012).
\bibitem{Mason} Cho, S. \textit{et al}. Symmetry Protected Josephson Supercurrents in Three-Dimensional Topological Insulators. \textit{Nature Comm.} $\mathbf{4}$, 1689 (2013).
\bibitem{Gordon} Williams, J. R. \textit{et al}. Unconventional Josephson Effect in Hybrid Superconductor-Topological Insulator Devices. \textit{Phys. Rev. Lett.} $\mathbf{109}$, 056803 (2012).
\bibitem{Dong_PRL} Xu, J.-P. \textit{et al}. Artificial Topological Superconductor by the Proximity Effect. \textit{Phys. Rev. Lett.} $\mathbf{112}$, 217001 (2014).




\bibitem{Leo} Mourik, V. \textit{et al}. Signatures of Majorana Fermions in Hybrid Superconductor-Semiconductor Nanowire Devices. \textit{Science} $\mathbf{336}$, 1003-1007 (2012).

\bibitem{Patrick Lee2} Liu, J. \textit{et al}. Zero-Bias Peaks in the Tunneling Conductance of Spin-Orbit-Coupled Superconducting Wires with and without Majorana End-States. \textit{Phys. Rev. Lett.} $\mathbf{109}$, 267002 (2012).
\bibitem{TeWari} Roy, D., Bondyopadhaya, N. \& Tewari, S. Topologically trivial zero-bias conductance peak in semiconductor Majorana wires from boundary effects. \textit{Phys. Rev. B} $\mathbf{88}$, 020502(R) (2013).
\bibitem{Marcus} Churchill, H. O. H. \textit{et al}. Superconductor-Nanowire Devices from Tunneling to the Multichannel Regime: Zero-Bias Oscillations and Magnetoconductance Crossover. \textit{Phys. Rev. B} $\mathbf{87}$, 241401(R) (2013).
\bibitem{Franceschi} Lee, E. J. H. \textit{et al}. Spin-resolved Andreev levels and parity crossings in hybrid superconductor-semiconductor nanostructures. \textit{Nature nanotech.} $\mathbf{9}$, 79-84 (2014).
\bibitem{Xue Nature physics QL} Zhang, Y. \textit{et al}. Crossover of the three-dimensional topological insulator Bi$_2$Se$_3$ to the two-dimensional limit. \textit{Nature Phys.} $\mathbf{6}$, 584-588 (2010).

\bibitem{Hedgehog} Xu, S.-Y. \textit{et al}. Hedgehog spin texture and Berry's phase tuning in a magnetic topological insulator.  \textit{Nature Phys.} $\mathbf{8}$, 616-622 (2012).
\bibitem{QL} Neupane, M. \textit{et al}. Observation of Quantum-Tunneling Modulated Spin Texture in Ultrathin Topological Insulator Bi$_2$Se$_3$ Films. \textit{Nature Commun.} $\mathbf{5}$, 4841 (2014).

\bibitem{BCS} Okazaki, K. \textit{et al}. Octet-Line Node Structure of Superconducting Order Parameter in KFe$_2$As$_2$. \textit{Science} $\mathbf{337}$, 1314-1317 (2012).



\bibitem{Point contact} Blonder, G. Tinkham, M. \& Klapwijk, T. Transition from metallic to tunneling regimes in superconducting microconstrictions: Excess current, charge imbalance, and supercurrent conversion. \textit{Phys. Rev. B} $\mathbf{25}$, 4515-4532 (1982).

\bibitem{Ag_Pb} van Son, P. C. van Kempen, H. \& Wyder, P. New method to study the proximity effect at the normal-metal-superconductor interface. \textit{Phys. Rev. Lett.} $\mathbf{59}$, 2226-2228 (1987).

\bibitem{Si_Nb} Heslinga, D. \textit{et al}. Observation of double-gap-edge Andreev reflection at Si/Nb interfaces by point-contact spectroscopy. \textit{Phys. Rev. B} $\mathbf{49}$, 10484-10494 (1994).



%

\bibitem{Zhou} Wang, E. \textit{et al}. Fully gapped topological surface states in Bi$_2$Se$_3$ films induced by a $d$-wave high-temperature superconductor. \textit{Nature physics} $\mathbf{9}$, 621-625 (2013).   
\bibitem{BSCCO_Hasan} Xu, S.-Y. \textit{et al}. Search for superconducting proximity effect in a topological insulator and high temperature superconductor heterostructure Bi$_2$Se$_3$/Bi$_2$Sr$_2$CaCu$_2$O$_{8+\delta}$. Preprint at http://arXiv:1403.2109 (2014).
\bibitem{BSCCO_Valla} Yilmaz, T. \textit{et al}., Absence of a Proximity Effect in a Topological Insulator on a Cuprate Superconductor: Bi$_2$Se$_3$/Bi$_2$Sr$_2$CaCu$_2$O$_8$. Preprint at http://arXiv:1403.4184 (2014).


\bibitem{NbSe2} Iwaya, K. \textit{et al}. Electronic state of NbSe$_2$ investigated by STM/STS. \textit{Physica B} $\mathbf{329-333}$, 1598-1599 (2003).

\bibitem{Rader} S\'anchez-Barriga, \textit{et al.}, J. Photoemission of Bi$_2$Se$_3$ with Circularly Polarized Light: Probe of Spin Polarization or Means for Spin Manipulation? \textit{Phys. Rev. X} $\mathbf{4}$, 011046 (2014).

\end{thebibliography}
\end{document}